 \documentclass[12pt,letterpaper]{article}	
\usepackage{linguex}	
\usepackage{times}
\usepackage{natbib}
 	\setcitestyle{semicolon,aysep={},yysep={,},notesep={; }}
\usepackage{lipsum}          
\usepackage{ragged2e}
\usepackage{graphicx}
\RaggedRight

\usepackage{scrextend}       
\deffootnote[.5em]{0em}{1em}{\textsuperscript{\thefootnotemark}\,}

\newcounter{savefootnote}   
\newcounter{symfootnote}
\newcommand{\symfootnote}[1]{%
   \setcounter{savefootnote}{\value{footnote}}%
   \setcounter{footnote}{\value{symfootnote}}%
   \ifnum\value{footnote}>8\setcounter{footnote}{0}\fi%
   \let\oldthefootnote=\thefootnote%
   \renewcommand{\thefootnote}{\fnsymbol{footnote}}%
   \footnote{#1}%
   \let\thefootnote=\oldthefootnote%
   \setcounter{symfootnote}{\value{footnote}}%
   \setcounter{footnote}{\value{savefootnote}}%
}

\usepackage[labelsep=period]{caption}

\usepackage[margin=1.0in]{geometry}
\usepackage[compact]{titlesec}
	\titleformat{\section}[runin]{\normalfont\bfseries}{\thesection.}{.5em}{}[.]
	\titleformat{\subsection}[runin]{\normalfont\scshape}{\thesubsection.}{.5em}{}[.]
	\titleformat{\subsubsection}[runin]{\normalfont\scshape}{\thesubsubsection.}{.5em}{}[.]
\usepackage[usenames,dvipsnames]{color}	
\usepackage[colorlinks,allcolors={black},urlcolor={blue}]{hyperref} 		

\renewenvironment{abstract}{%
\noindent\begin{minipage}{1\textwidth}
\setlength{\leftskip}{0.4in}
\setlength{\rightskip}{0.4in}
\textbf{Abstract.}}
{\end{minipage}}
\newenvironment{keywords}{%
\vspace{.5em}
\noindent\begin{minipage}{1\textwidth}
\setlength{\leftskip}{0.4in}
\setlength{\rightskip}{0.4in}
\textbf{Keywords.}}
{\end{minipage}}

\begin{document} 

\setlength{\Extopsep}{6pt}
\setlength{\Exlabelsep}{9pt}		

\begin{center}			
\normalfont\bfseries
Human-AI Interactions Through A Gricean Lens
\vskip .5em
\normalfont
{Laura Panfili, Steve Duman, Andrew Nave, Katherine Phelps Ridgeway, Nathan Eversole \& Ruhi Sarikaya\symfootnote{Authors: Laura Panfili, Amazon (\href{mailto:lpanfili@amazon.com}{lpanfili@amazon.com}), Steve Duman, Amazon (\href{mailto:dumastev@amazon.com}{dumastev@amazon.com}), Andrew Nave, Amazon (\href{mailto:andrnave@amazon.com}{andrnave@amazon.com}), Katherine Phelps Ridgeway (\href{mailto:k1000ridgeway@gmail.com}{k1000ridgeway@gmail.com}), Nathan Eversole, Amazon (\href{mailto:neeverso@amazon.com}{neeverso@amazon.com}) \& Ruhi Sarikaya, Amazon (\href{mailto:rsarikay@amazon.com}{rsarikay@amazon.com})
.}}
\vskip .5em
\end{center}

\begin{abstract}
Grice’s Cooperative Principle (\citeyear{grice_1975}), which describes the implicit maxims that guide effective conversation, has long been applied to conversations between humans. However, as humans begin to interact with non-human dialogue systems more frequently and in a broader scope, an important question emerges: what principles govern those interactions? In the present study, this question is addressed, as human-AI interactions are categorized using Grice’s four maxims. In doing so, it demonstrates the advantages and shortcomings of such an approach, ultimately demonstrating that humans do, indeed, apply these maxims to interactions with AI, even making explicit references to the AI’s performance through a Gricean lens. Twenty-three participants interacted with an American English-speaking Alexa and rated and discussed their experience with an in-lab researcher. Researchers then reviewed each exchange, identifying those that might relate to Grice’s maxims: Quantity, Quality, Manner, and Relevance. Many instances of explicit user frustration stemmed from violations of Grice’s maxims. Quantity violations were noted for too little but not too much information, while Quality violations were rare, indicating high trust in Alexa’s responses. Manner violations focused on speed and humanness. Relevance violations were the most frequent of all violations, and they appear to be the most frustrating. While the maxims help describe many of the issues participants encountered with Alexa’s responses, other issues do not fit neatly into Grice’s framework. For example, participants were particularly averse to Alexa initiating exchanges or making unsolicited suggestions. To address this gap, we propose the addition of human \textbf{Priority} to describe human-AI interaction. Humans and AIs are not (yet?) conversational equals, and human initiative takes priority. Moreover, we find that Relevance is of particular importance in human-AI interactions and suggest that the application of Grice’s Cooperative Principles to human-AI interactions is beneficial both from an AI development perspective as well as a tool for describing an emerging form of interaction.
\end{abstract}

\begin{keywords} 
discourse; pragmatics; Cooperative Principle; artificial intelligence; Alexa; voice assistant
\end{keywords}

\section{Introduction}

Grice’s (\citeyear{grice_1975}) Cooperative Principle provides a universal framework of how humans negotiate the intricate practice of conversation. Grice developed this approach to describe successful and unsuccessful human-human interactions. However, in recent years, humans are increasingly engaged in language-based interaction with a new type of interlocutor: voice-based Artificial Intelligence (AI). From phones to at-home speakers to in-car devices, AIs — as manifest in voice assistants like Alexa, Google Assistant, Siri, and Cortana — are pervasive in a myriad of contexts. The ubiquity of these devices, coupled with steady improvement of the models that drive them, lead to two challenges. First, developers of AI technology need mechanisms to evaluate the success or failure of exchanges with AI. Second, researchers require a method by which to describe the principles that guide this new, pervasive form of language-based exchanges. 
 
The study we report here takes the novel step of applying Grice’s framework, one meant to describe human conversation, to exchanges that challenge the notion of what constitutes a ‘conversation.’ We rely largely on qualitative observations of in-lab, naturalistic exchanges between participants and an Alexa device. Participants rated Alexa’s responses and were prompted by an in-lab researcher to describe the reasoning behind their ratings. The researchers then reviewed all exchanges, identifying over 700 as potential violations of Grice’s maxims. In the results reported here, we show the exchanges, ratings, and explanations from this study. These observations provide strong preliminary evidence that users do, indeed, apply Gricean maxims to human-AI interactions.
 
Based on our findings, we argue that Grice’s Cooperative Principle — specifically the maxims of Quantity, Quality, Relevance, and Manner — is a useful framework to describe human-AI interactions, though there are a few important differences from what we expect with human-human interactions. With regard to Quantity, participants specifically cite responses that are too short as reasons for providing low ratings of Alexa’s responses, but they do not explicitly mention lengthy responses. Quality violations are rare in the dataset, with very few explicit references by participants as to the quality of Alexa’s responses. Manner violations are observed with regard to the rate of Alexa’s speech and the ‘humanness’ of responses. Violations of Relevance are the most prevalent violations in the dataset, and they are most frequently associated with lower ratings.
 
In addition, we also identify another important difference between human-human interactions and those displayed in human-AI interactions: the inherent power difference between humans and AIs leads to different expectations in terms of turn initiations and interruptions. To account for the discrepancies, we propose a new maxim, Priority, which suggests that humans take priority in human-AI interactions.
 
To explore these observations, we first examine previous literature regarding Grice’s Cooperative Principle, criticisms regarding its universality, and applications regarding differing discourse contexts. Next, we present the study of twenty-three participants interacting with Alexa in a lab setting, focusing primarily on the qualitative observations of the participants. Lastly, we discuss our findings and their implications regarding AI evaluation, the limitations of Grice’s framework, and future directions.

\section{Background}

Understanding user perceptions of AI systems is notoriously difficult, but it is an important endeavor. Evaluating the performance of AI systems allows us to identify problem areas as the technology develops. Specifically, how do human users assess the success or failure of their exchanges with an AI dialogue system? While several approaches have been suggested for doing so (see, e.g., \citealt{deriu_etal_2019, schmitt_ultes_2015}), previous attempts are limited in their scope, examining voice user interfaces that are primarily transactional or limited to a particular domain, like setting up travel plans. These limitations become apparent when examining AI voice assistants like Siri, Google Assistant, and Alexa, which now feature numerous functions (including naturalistic conversation, in some cases) and can be used in a variety of contexts. Moreover, the machine learning models that drive the Natural Language Understanding of these systems are steadily improving, allowing users to employ more complex linguistic constructions. As a consequence, it is important to have a well-established, human-based approach with which to understand how humans interpret their own interactions with AI systems. 
 
In addition, as humans interact with AIs more frequently and more naturally, there is increasing evidence that users anthropomorphize AIs, talking and perhaps even treating them as human-like entities. \cite{purington_2017} demonstrate that Alexa users frequently personify the device in reviews posted to Amazon.com (e.g., using female pronouns ‘she’ and ‘her’ to describe Alexa), and, interestingly, this personification is correlated with higher star ratings. Given the increasingly complex interactions along with the natural tendency (and even encouragement) to personify AIs, it makes sense to apply frameworks traditionally used for human-human interactions in order to understand people’s perceptions of human-AI interactions. To that end, in this paper we explore the utility of using Grice’s Cooperative Principle — a framework traditionally used to describe interactions between humans — as a framework with which to understand user perceptions of interactions between humans and AIs.
 
Grice’s Cooperative Principle is a descriptive framework to explain the presumably universal guidelines for effective conversations. The principle is composed of four maxims: Quantity, Quality, Manner, and Relevance. These maxims are intended to describe the implicit conversational rules that allow humans to mutually understand one another. These maxims are briefly described as follows:\\

\medskip

\begin{addmargin}[1em]{2em}

\noindent\textit{\textbf{Quantity:} Make your contribution as informative as is required for the current purposes of the exchange. Do not make your contribution more informative than required.}

\noindent\textit{\textbf{Quality:} Try to make your contribution one that is true, specifically: Do not say what you believe to be false; Do not say that for which you lack adequate evidence}
 
\noindent\textit{\textbf{Manner:} Be perspicuous, and specifically: Avoid obscurity; Avoid ambiguity; Be brief; Be orderly}
 
\noindent\textit{\textbf{Relevance:} Make your contributions relevant}

\end{addmargin}

\medskip
 
In order for human conversation to be effective, humans need to follow Grice’s maxims. Violating the maxims leads to ineffective (or, in the case of flouting, often comical) communicative events.
 
Since Grice’s seminal work, many researchers have argued that the Cooperative Principle is not necessarily universal, as contextual factors can influence the extent to which interlocutors follow the maxims \citep{habermas_1984, harris_1995, thornborrow_2014}. As Harris (1995) argues, “Grice has often been criticised for constructing a theory of universal pragmatics which cannot handle unequal encounters which are clearly not cooperative and where the goals of participants conflict in quite obvious ways.” Such situations include ‘institutional discourse,’ or discourse that takes place in institutions like classrooms or courtrooms. In a courtroom setting, for example, a witness on the stand does not have the access to the same speech acts as the lawyer or judge \citep{habermas_1984}. 

Kitis (\citeyear{kitis_1999}) addresses the universality of Grice’s Cooperative Principle by suggesting that, in fact, Relevance serves as a ‘supermaxim’ that structures surrounding communication and exerts control of the pragmatic definition of truth. Relevance, Kitis argues, thus constrains the application of the other three maxims. This accounts for power dynamics in human-human interactions in that the interlocutors respond according to the ‘relevance’ of the social and linguistic context. This line of thinking suggests that one maxim can influence or filter the others. Like in human-human conversation, the context of human-AI interactions can vary wildly (e.g., in the kitchen, in the car, etc.). While the notion of ‘institutional discourse’ covers many of the contextual variations in Relevance, it does not necessarily capture the power differences at play between human and AI. This point will be taken up in the Discussion below.
 
In short, despite limitations in Grice’s Cooperative Principle for describing human conversation, it is a well-established and useful starting point when endeavoring to explore human perceptions of human-AI interactions. Moreover, applying this framework could help us improve the development process of AIs, help us better categorize and understand human-AI exchanges, and perhaps even contribute our understanding of Grice’s maxims. 
 
To these ends, this paper describes an in-lab experiment in which we examine human-AI exchanges through a Gricean lens. As this is a novel approach to such interactions, our methodology is primarily qualitative, where we analyze human-AI exchanges through three means: (1) the researchers’ assessments of human-AI interactions as violations of Gricean’s maxims, (2) participants’ assessments of exchanges on a 1-5 scale, and (3) participants’ explicit references to exchanges that the researchers regard as applicable to Grice’s maxims. This methodology allows us to determine the extent to which applying Grice’s Cooperative Principle to human-AI interactions is useful as well as the extent to which participants consciously (and perhaps unconsciously) do so as well. More specifically, it allows us to address the following research questions: \\

\begin{addmargin}[1em]{2em}

\medskip

\noindent{\textbf{RQ1:} Are unsatisfactory interactions between humans and AIs explained as violations of Grice’s maxims?}\\
\noindent{\textbf{RQ2:} Which, if any, unsatisfactory interactions are not explained as violations of Grice’s maxims?} \\
\noindent{\textbf{RQ3:} Are there fundamental differences between what researchers perceive to be violations of Grice’s maxims compared to participant assessments?} \\

\medskip

\end{addmargin}
 
In order to answer these research questions, we invited twenty-three participants to a lab where they interacted with an Alexa device. Alexa was chosen first and foremost because this study was conducted by researchers at Amazon, Inc. to help facilitate the development of Alexa. Additionally, Alexa is a widely used voice assistant with a wide range of capabilities. The details of the methodology are reported below.

\section{Methods}

\begin{figure}[htp]

\centering
\includegraphics[width=.3\textwidth]{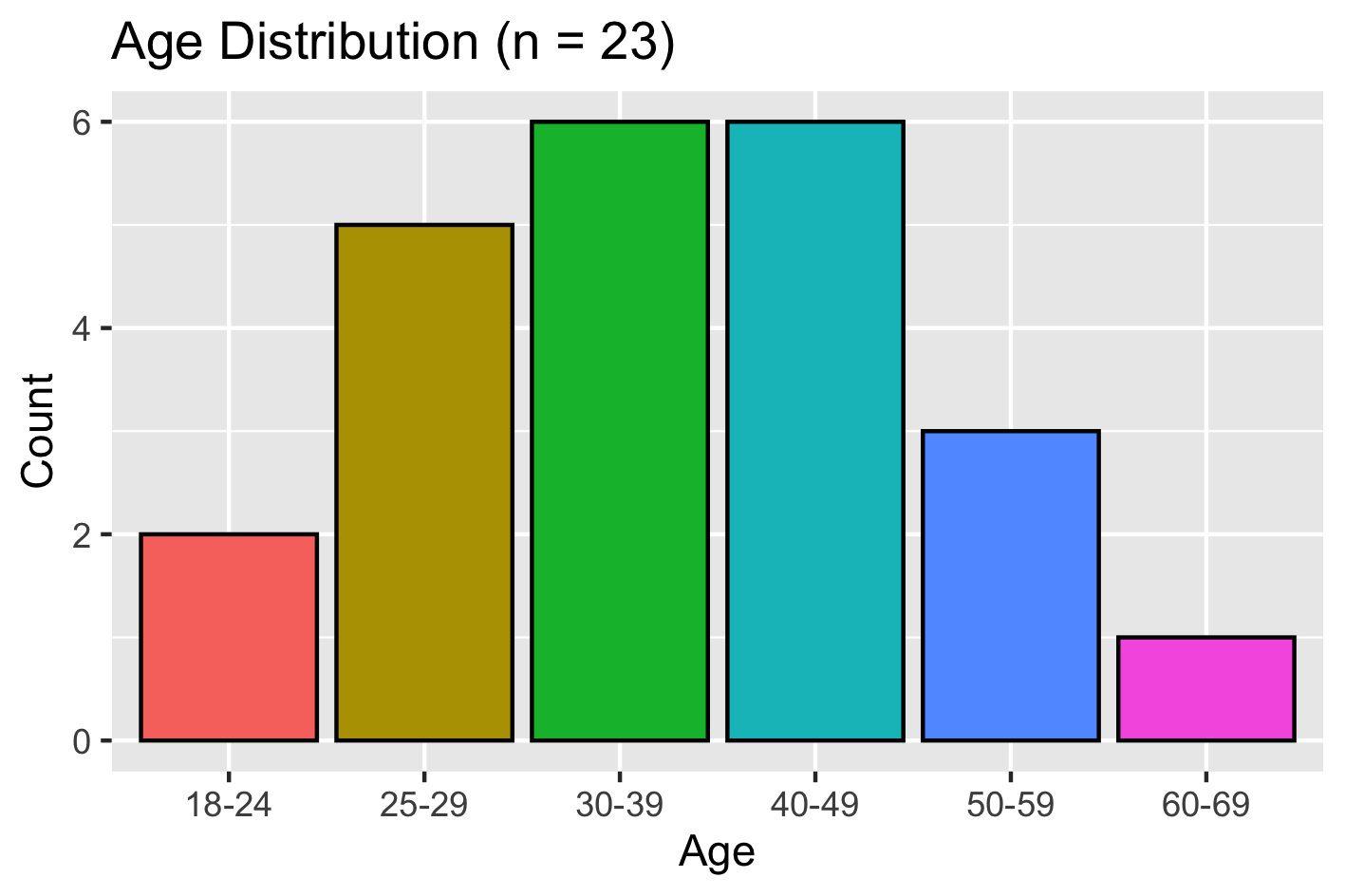}\hfill
\includegraphics[width=.3\textwidth]{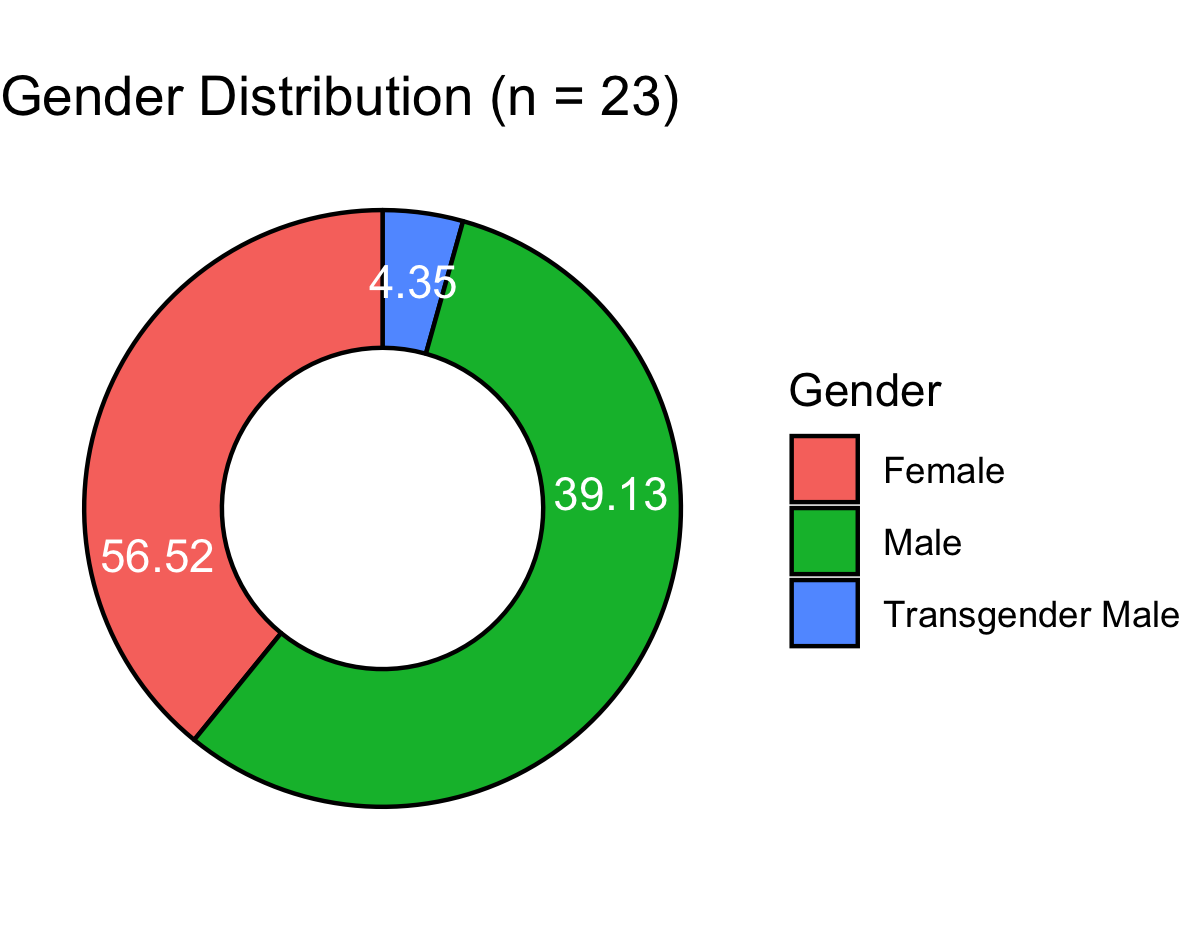}\hfill
\includegraphics[width=.3\textwidth]{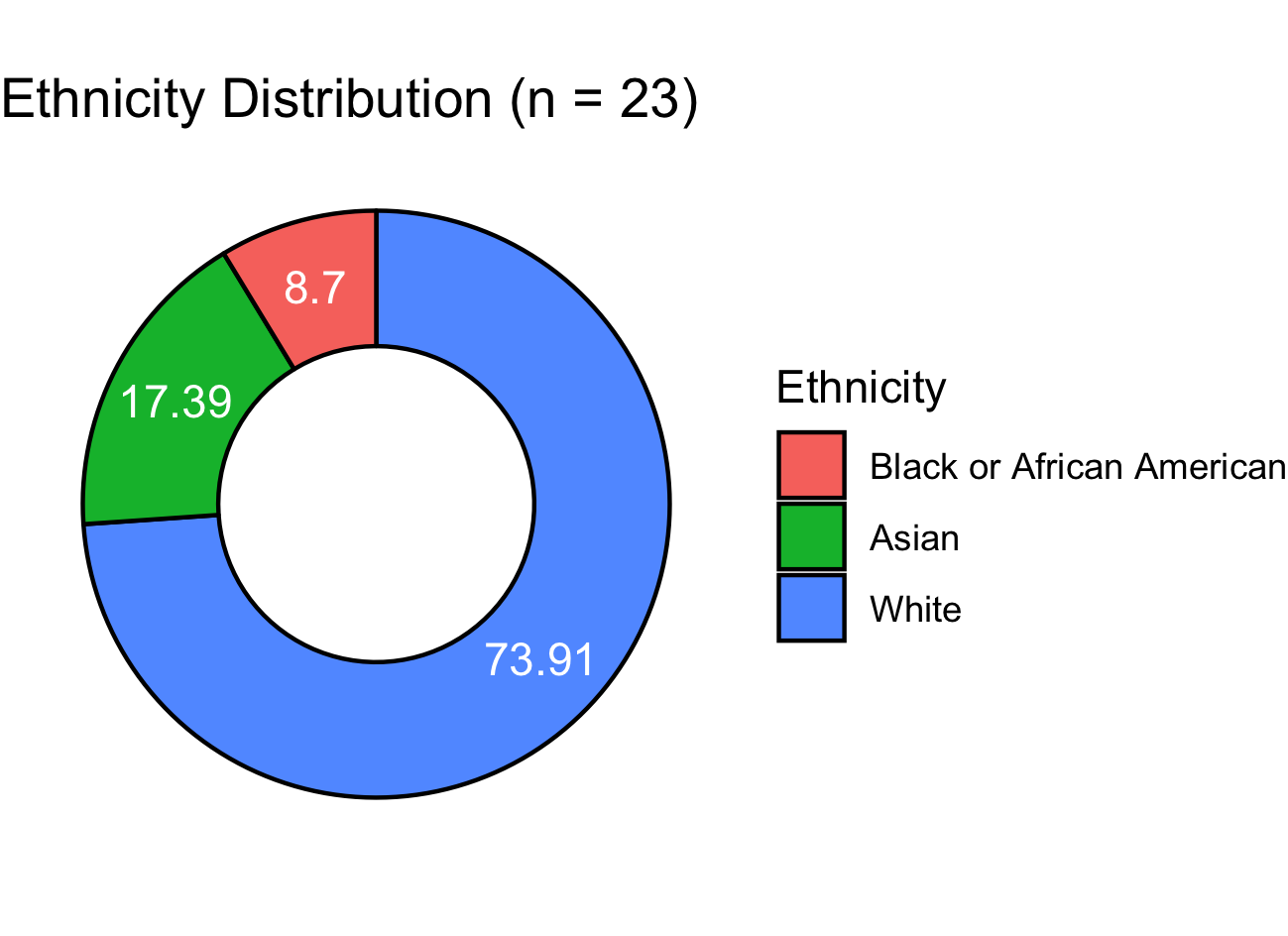}

\caption{Age, gender, and ethnicity of participants}
\label{fig:demographics}

\end{figure}

Amazon researchers worked with User Research International to recruit twenty-three participants for this study. Participants were self-reported to have an Alexa-enabled device at home and to interact with it at least a few times each week. All participants reported that they had never professionally worked on a voice assistant and did not have any immediate family members who did, nor were any employed by Amazon. Participants were all residents of the greater Seattle area and able to speak and read English, though not necessarily natively. They received a \$100 Amazon.com gift card in exchange for participating in this experiment. Figure \ref{fig:demographics} shows participant demographics.

The study was conducted by Amazon researchers in a usability lab on Amazon’s campus in Seattle, Washington in the fall of 2019. This lab was set up with a third generation Echo Dot set to American English and connected to various Alexa-enabled smart home devices. An in-lab researcher joined each participant in order to troubleshoot technical issues, facilitate ratings, and encourage discussion of their reasoning behind ratings.
 
Participants were instructed to interact with Alexa based on a series of prompts; they were encouraged to explore anything that piqued their interest and were free to move on to the next prompt at any point. The prompts were designed to cover some of the most frequent types of Alexa exchanges, such as music and smart home requests. They were intended to cause both successful and unsuccessful interactions and to allow participants to phrase their requests naturally. Prompts were displayed on a monitor and participants controlled the pace with a keyboard. Audio of each session was recorded for later analysis. Prompts were presented in a random order, with not all participants completing all prompts, depending on their pace.
 
Participants were asked to rate each of Alexa’s responses an a scale from one to five, where one is the worst score and five is the best. The points on the scale were not further specified, allowing for the participants to apply the scale as they felt was intuitive. Scores were input using number keys on a modified keyboard and recorded using Psychopy software \citep{peirce_etal_2019}. In order to preserve the naturalness of the experiment, participants were only lightly encouraged and reminded to rate their exchanges, resulting in some unrated turns. Additionally, the in-lab researcher frequently asked participants to talk through their thought process behind a rating. At the end of the experiment, participants verbally completed an exit survey that included a broad discussion of their application of the rating scale.
 
\begin{figure}
  \centering
  \includegraphics[width=0.5\linewidth]{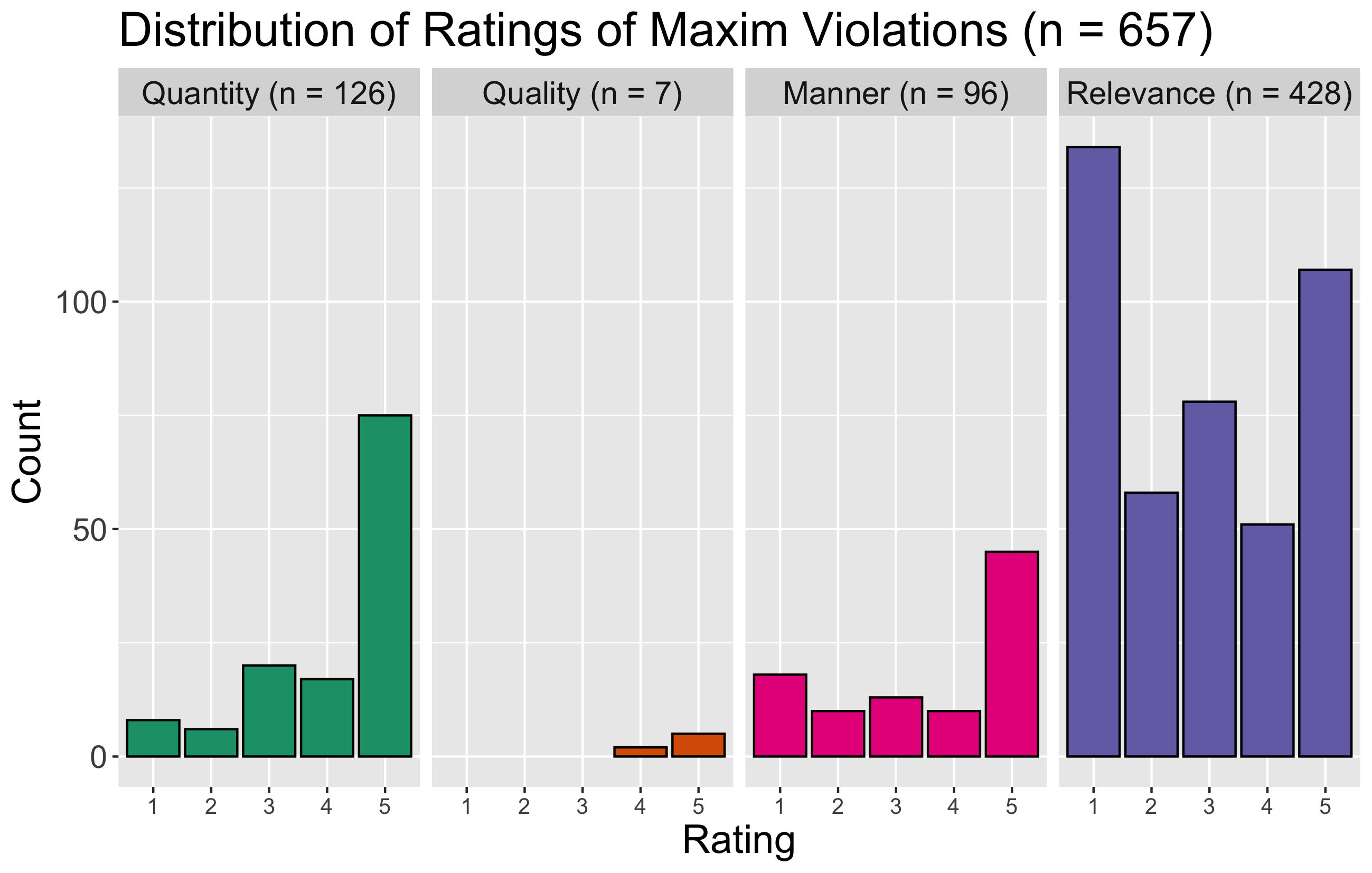}
  \caption{Distribution of ratings of maxim violations}
  \label{fig:ratings_all}
\end{figure}

The resulting data set consists of 23 hours of audio, including exchanges with Alexa and discussions with the in-lab researcher. We captured 2,279 participant-Alexa exchanges, 1,913 of which were rated on the 1-5 scale. All exchanges were evaluated by researchers for potential violations of one or more maxim on Alexa’s part; we also identified implicit and explicit references to Grice’s maxims in participant conversations with the researchers. Table \ref{tab:counts} shows the frequency of maxim violations and Figure \ref{fig:ratings_all} shows the distribution of ratings by maxim violation. 

\begin{table}[h]
\centering
	\begin{tabular}{l c}
	Maxim & n\\
	\hline
	Quantity &  135 \\
	Quality & 7 \\
	Manner & 109 \\
	Relevance & 471 \\
	\hline
	\end{tabular}
\caption{Maxim violation counts \label{tab:counts}}
\end{table}

As the table demonstrates, approximately 40\% of the participant-Alexa exchanges were identified the researchers as potential Gricean violations. The most frequently violated maxim was Relevance, where 428 exchanges were tagged as potential violations. Interestingly, as Figure \ref{fig:ratings_all} demonstrates, violations of Grice’s maxims are associated with both negative and positively-rated turns. In other words, a researcher-identified maxim violation is not necessarily a poorly rated exchange by the participants. In fact, many researcher-identified violations are rated 5 by participants. However, this seems to be less the case for Relevance, where the distribution of ratings includes much lower numbers. The following section explores the findings for each maxim, focusing primarily on participants references to Grice’s maxims in their evaluations.

\section{Results}

This section consists of a primarily qualitative analysis of the participant-Alexa exchanges recorded in the study. The subsections discuss exchanges that the researchers identified as possible violations of Grice’s four maxims: Quantity, Quality, Manner, and Relevance. In addition to the participant ratings for each exchange, we concentrate primarily on explanations that either implicitly or explicitly reference the maxim. 

\subsection{Quantity}

We identified 135 exchanges where Alexa may have violated the Quantity maxim, with an average rating of 4.15 (SD = 1.22). We saw several trends in how participants invoked Quantity in their conversations with the in-lab researcher. First, some participants noted and appreciated when Alexa’s response hit the Quantity sweet spot (such examples are not included as violations, but rather demonstrate participants explicitly referring to Gricean maxims), as seen in Example \ref{ex:thomas_cook}.

\ex. \label{ex:thomas_cook} 
\textbf{User:} Alexa, what's happening with the Thomas Cook collapse over in Europe?\\
\textbf{Alexa:} Here's what I found about Thomas Cook from Reuters. The collapse of British travel operator Thomas Cook has left hundreds of thousands holidaymakers abroad and forced governments and insurers to coordinate a huge operation to get them home. The company ran hotels, resorts and airlines ferrying 19 million people a year to 16 different countries. \textit{[rating: 5]\\}
\textbf{User:} The thing I liked about that is it wasn’t too long. So I got a good amount of info […] It was short and sweet.\\
 
When Alexa’s response did not hit the sweet spot and instead provided too little information, this quantity violation was sometimes cause for a lower rating. In Example \ref{ex:bananas}, Alexa seemed about to provide a list, but only included one item.
 
 \ex. \label{ex:bananas} 
\textbf{User:} Alexa, tell me about different kinds of bananas. \\
\textbf{Alexa:} Some examples of bananas are red bananas. \textit{[rating: 2]} \\
\textbf{User:} There should’ve been more. I’m giving that a two. I wanted more, like … yellow and plantains or whatever, I wanted three or four. \\
 
However, when Alexa provided what would likely be too much information in a conversation between humans, this generally went unnoticed by participants. In Example \ref{ex:astronomy}, Alexa read the entire first paragraph of the Wikipedia article on Astronomy. Though the device does ask if the user would like to hear more, it’s only offered after reading that full first paragraph. Another potential quantity violation comes in the form of too much detail: in Example \ref{ex:paris}, Alexa provides distances down to the tenth of a mile for locations on another continent, and the participant did not comment on this. 
 
 \ex. \label{ex:astronomy} 
\textbf{User:} Alexa, tell me about astronomy.\\
\textbf{Alexa:} Here's the Wikipedia article on Astronomy. Astronomy is a natural science that studies celestial objects and phenomena. It applies mathematics, physics, and chemistry in an effort to explain the origin of those objects and phenomena and their evolution. Objects of interest include [...] Would you like me to continue? \textit{[rating: 5]}\\
 
 \ex. \label{ex:paris} 
\textbf{User:} Alexa, tell me about good restaurants in Paris, France.\\
\textbf{Alexa:} Here are a few top rated restaurants in Paris, France: one, Chez le Libanais, 4,997.6 miles away on et 36 rue Saint-André-des-Arts, two, La Jacobine, 4,997.5 miles away on 59-61 rue Saint-André des Arts [...] \textit{[rating: 4]}\\
 
These examples demonstrate that some users seem to have very different expectations of Alexa compared to humans in terms of an adherence to the maxim of Quantity. Specifically, these users seem to find responses that we, as researchers, interpret as too much information — such as an entire paragraph from Wikipedia or distances in thousands of miles to the decimal point — as perfectly acceptable. This suggests that some users are sensitive to too little quantity, but they are less sensitive to greater quantity, in contrast to what we may expect with human-human interactions. This marks a first potential difference in the application of Grice’s framework to human-AI interactions.
 
These examples may be a reflection of users viewing the AI as a window to the Internet. Many participants demonstrated a belief that Alexa is conducting Internet searches to answer their questions, and that the device can access whatever information an Internet search could access. If Alexa is capable of accessing all the information that’s out there, responses should provide more detail than a human can; on the other hand, this expectation sets users up for disappointment and frustration when Alexa fails to provide sufficient information.

\subsection{Quality}

Just seven exchanges included Alexa violations of the Quality maxim, with an average rating of 4.71 (SD = 0.49). Participants demonstrated overwhelming trust in Alexa’s responses, as indicated by high ratings and commentary to that effect along with the relative lack of explicit references to Quality. In Example \ref{ex:mordor}, the participant was asked to request the weather in Mordor, a fictional location from The Lord of the Rings, but this participant did not know that Mordor does not exist. Due to a speech recognition error, Alexa reported the weather for Vidor, Texas instead. The participant then assumed that Alexa was correctly reporting the weather in Mordor, Texas:
 
\ex. \label{ex:mordor} 
\textbf{User:} Alexa, what’s the weather like in Mordor?\\
\textbf{Alexa:} In Vidor, Texas, it's 90 degrees Fahrenheit with mostly cloudy skies. Today, you can look for partly sunny weather, with a high of 90 degrees and a low of 77 degrees. \textit{[rating: 5]}\\
\textbf{User:} Is that in Texas? I’m gonna say she’s right. \\
 
In another example \ref{ex:rainier}, Alexa gave factually incorrect information, which did not seem to bother the participant: 
 
\ex. \label{ex:rainier} 
\textbf{User:} Alexa, what's the highest point in the Seattle area?\\
\textbf{Alexa:} The highest point in Seattle is Mount Rainier. \textit{[rating: 5]}\\
 
Mount Rainier is about a three-hour drive from Seattle, not in Seattle (a fact we can assume any Seattle area resident knows). Nonetheless, the participant rated this response a five. This is perhaps complicated by the fact that the original request was, in fact, about the Seattle area, but this factually incorrect answer did not seem to cause any broad mistrust in Alexa’s responses.
 
One of the more striking examples of trust in Alexa comes from a participant who, after several failed attempts to find out who holds the home run record for the Seattle Mariners, concluded that that information doesn’t exist; she has so much faith in Alexa’s ability to find information that she interpreted Alexa’s repeated failures as an exhaustive search.
 
\ex. \label{ex:mariners} 
\textbf{User:} Who hit the most home runs for the Seattle Mariners?\\
\textbf{Alexa:} The Houston Astros at Seattle Mariners game hasn't started yet. First pitch will be tomorrow evening at 7:10pm. \textit{[rating: 1]}\\
\textbf{User:} Which player on the Seattle Mariners has hit the most home runs?\\
\textbf{Alexa:} Hmm, I'm not sure. \textit{[rating: 3]}\\
\textbf{User:} Who has the home run record for Seattle Mariners?\\
\textbf{Alexa:} This might answer your question: Yesterday, the Mariners did not hit a home run during their 2-1 loss to the Orioles. This season, the Mariners are ranked 5th in the American League with 238 home runs. This season, the Mariners are 66 and 90 and ranked 5th in the AL West. Did that answer your question? \textit{[rating: 4]}\\
\textbf{User:} [...] The second time around she attempted to answer it, and then was like 'wait, is that really what you were asking for?' and asking that was sort of like, I’m now understanding that maybe that stat’s not out there in the universe. [rating: 4]\\
 
Alexa occasionally cites sources, and we saw just one instance of a participant questioning Quality in the form of doubting a source. This participant rated the exchange a five but expressed some disappointment with the source of the information. (This participant explicitly refused to rate any response below a three.)
 
 \ex. \label{ex:giraffe} 
\textbf{User:} Alexa, how many spots does a giraffe have?\\
\textbf{Alexa:} According to another Amazon customer: no individual giraffe has the same pattern so there is not a way to tell how many spots a giraffe has. Did that answer your question? [rating: 5]\\
\textbf{User:} No, who is this other Amazon customer?\\
 
In sum, participants in the study demonstrated a very strong trust in Quality from Alexa, accepting most of the information that was provided, even in cases where Alexa provided factually incorrect information or did not provide information that definitely exists. This could suggest that, because Alexa is an AI entity, humans interlocutors have even higher expectations of Quality than they do for humans. Like Quantity, this evaluation of Quality may be impacted by the perception of Alexa as a window to the Internet. We saw participants generally trust the information that Alexa provides, as they likely would trust the results of an Internet search (whether or not they should trust an Internet search is another topic, of course). Judgement of Quality seems to parallel an Internet search: It is hard to believe that an Internet search would fail to find information that does exist, and the trustworthiness of sources can be questioned. Users seem to pin these failures not on Alexa but rather on the Internet search that they believe the device conducts.

\subsection{Manner}

109 exchanges were tagged as including potential Manner violations on Alexa’s part, with an average rating of 3.56 (SD = 1.59). Several participants noted that Alexa spoke too quickly, posing comprehension problems. Though not mentioned by participants, unnatural prosody may have compounded the difficulties posed by speed. Examples \ref{ex:skills} and \ref{ex:thermostat} show two participant reactions to fast Alexa responses.
 
\ex. \label{ex:skills} 
\textbf{User:} Do you have any financial skills?\\
\textbf{Alexa:} Here are three popular Business \& Finance skills. First is Make Me Smart. Say `Alexa, Make Me Smart' to learn something surprising and timely about the economy, pop culture, and tech from Marketplace’s Kai Ryssdal and Molly Wood. Would you like to give that one a try? \textit{[rating: 3]}\\
\textbf{User:} No.\\
\textbf{Alexa:} Okay, next is [...] Would you like to give that one a try? \textit{[rating: 3]}\\
\textbf{User:} No.\\
\textbf{Alexa:} Okay, next is [...] Would you like to give it a try? \textit{[rating: 3]}\\
\textbf{User:} No.\\
\textbf{Alexa:} Okay. Do you want to hear more? \textit{[rating: 5]}\\
\textbf{User:} No.\\
\textbf{Alexa:} Alright. \textit{[rating: 5]}s\\
\textbf{User:} It’s just so fast, it’s like, oh gosh! [...] a lot, and a little quick. Not that I think there's a way to change the speed of the speech, but it seemed really fast.\\
 
\ex. \label{ex:thermostat} 
\textbf{User:} Alexa, adjust the thermostat to sixty five degrees.\\
\textbf{Alexa:} Auto's keeping the temperature between 32 and 122.\\
\textbf{User:} Alexa, can you adjust the thermostat to sixty five degrees.\\
\textbf{Alexa:} Auto's keeping the temperature between 32 and 122. \textit{[rating: 3]}\\
\textbf{User:} I don’t really know what Alexa was saying. Alexa was reading it kinda fast. Not even the second time, I didn’t catch what Alexa was saying. \\
 
Just one participant commented on obscurity of language, suggesting that the word ‘intermittent’ was likely not accessible to all audiences.
 
 \ex. \label{ex:intermittent} 
\textbf{User:} Alexa, what’s the weather like today?\\
\textbf{Alexa:} Currently, in Seattle, it’s 65 degrees Fahrenheit with partly sunny skies. Today, you can expect intermittent clouds with a high of 67 degrees and a low of 59 degrees. \textit{[rating: 4]}\\
\textbf{User:} It was good. I don’t know if people would understand what the word ‘intermittent’ means, that’s all.\\
 
Alexa regularly provided rather clunky responses — at least from the human-human conversation perspective — that went unmentioned by participants. In Example \ref{ex:music}, Alexa provides a lot of information in response to a request for music. This would likely feel unnatural in a conversation between friends, but Alexa’s status as a non-human interlocutor may impact evaluation.
 
 \ex. \label{ex:music} 
\textbf{User:} Alexa, play some country music.\\
\textbf{Alexa:} Okay. The station: Top Country, free on Amazon Music. \textit{[rating: 5]}\\
 
On the whole, violations of Manner are relatively infrequent in the dataset, and those that are present—references to speaking speed or vocabulary—are what we might expect from human-human interactions. However, one difference from our expectations did appear in the dataset. Despite many participants overlooking unnatural phrasing, we saw a few hints that humanness may be an important aspect of Manner when interacting with Alexa. For example, one participant noted, “The ones that I gave a [...] three were the ones that were more … how do I say, it was like mechanical or rote.” Evaluation of humanness likely varies from person to person, depending on what expectations users have of their AI’s capabilities and roles. Regardless, this would be an important difference from expectations of human-human interactions, where AIs are judged in terms of how ‘human’ their responses are.

\subsection{Relevance}

Finally, we identified 471 exchanges where Alexa may have violated the Relevance maxim, with an average rating of 2.86 (SD = 1.58), making Relevance the most frequently-violated maxim and that with the lowest average rating compared to the other maxims (a pair of Welch two-sample t-tests demonstrates that its mean is significantly lower than that of Quantity (Mean = 4.15, SD = 1.23, t = 9.7, df = 259, p $<$ .001) and Manner (Mean = 3.56, SD = 1.59, t = 3.9, df = 140, p $<$ .001))\footnote{There are too few observations from the Manner maxim to run a valid statistical comparison with Relevance.}.
 
Two broad types of Relevance violations emerged in our data set. The first is \textbf{irrelevant arguments}, where a key argument, such as a name or a place, is different in the user’s request and Alexa’s response. In Example \ref{ex:monte_carlo}, the participant asks about information for one location, and Alexa responds with that information for a different location.
 
  \ex. \label{ex:monte_carlo} 
\textbf{User:} Alexa, what language do they speak in Monte Carlo, Monaco? \\
\textbf{Alexa:} This might answer your question: Nossa Senhora do Monte is in Portugal, where the official languages are Portuguese and Mirandese. \textit{[rating: 2]}
 
The second type of Relevance violation we found is \textbf{irrelevant content}, where the goal, rather than the arguments, is irrelevant. Example \ref{ex:thermostat_name} shows an instance of the participant asking for a smart home device name and getting different information about that device.
 
   \ex. \label{ex:thermostat_name} 
\textbf{User:} Alexa, what did I name my thermostat?\\
\textbf{Alexa:}  The thermostat temperature is 68 degrees. \textit{[rating: 3]}\\
\textbf{User:} I’m gonna give it a three because that isn’t really what I asked for.\\
 
When participants were asked to describe reasons that they would provide a particular rating, they cited Relevance much more frequently than any of the other three maxims. Participants described low ratings as including Relevance violations, noting that “One was where she told me something that I didn’t even want to know and it went in the wrong direction” and “A two would probably be asking what’s the weather in Seattle and it giving me Portland.”
 
These findings show that violations of Relevance are the most frequent and salient Gricean violations in interactions with Alexa. Relevance, which has previously been proposed as a supermaxim \citep{kitis_1999}, seems to have a different status from other maxims in human-AI interactions, with more violations and a lower average rating. The higher occurrence of Relevance violations likely speaks more to Alexa’s responses than to how humans apply Grice’s maxims: Relevance violations can be the result of a failure at effectively any step of the complex pipeline powering voice AIs, such as speech recognition errors or natural language understanding errors. However, the lower ratings indicate that human interlocutors find these violations particularly grating. Some of this may stem from the largely transactional nature of human-AI interactions, as Relevance violations impede the basic expected function of the device.

\section{Discussion}

Overall, the above findings indicate that Grice’s Cooperative Principle is a useful framework with which to describe interlocutor assessments of human-AI interactions. Specifically, we can make the following high-level observations for each maxim:
 
With regard to \textbf{Quantity}, participants explicitly disapprove of exchanges where Alexa provided less information than they expected. Interestingly, however, in contrast to what the researchers would likely expect from an equivalent human-human interaction, participants do not explicitly disapprove of several exchanges where Alexa provided arguably too much information.
 
For \textbf{Quality}, participants in the study demonstrate a significant amount of trust in Alexa’s responses. In fact, one participant was so trusting that Alexa’s inability to provide a statistic led her to believe that the statistic did not exist. This extreme trust in Alexa’s responses is likely driven both by the assumption that Alexa is following the maxim as well as lay-person assumptions about AI being ‘all-knowing.’ This is a particularly interesting finding that warrants extensive research, for, as AI becomes more prolific, people will likely more and more frequently use AI as a knowledge resource. Since no system (or person) will ever be 100\% accurate, it is important that participants have a measured sense of Quality in their interactions with AIs.
 
In terms of \textbf{Manner}, participants make explicit reference to the speed at which Alexa speaks (likely a way of talking about prosody as well), and one even mentioned obscurity as a potential problem. User perceptions of Alexa’s humanness may play a role in their perception of Manner. Interestingly — perhaps related to Quantity — the amount that Alexa reports when performing an action like playing a song was not explicitly referenced as problematic. 

In this study, \textbf{Relevance} is far and away the most commonly violated maxim as well as the mostly lowly rated on average. It is also the most explicitly mentioned by participants in the study. Participants cited relevance (worded as “the wrong direction,” not “what I asked for,” etc.) as a primary reason for providing a low rating in exchanges with Alexa. Relevance violations can be particularly detrimental to more transactional exchanges, which are common for human-AI interactions.

\subsection{The Priority Maxim}

In addition to the observations of the above maxims, we also observed an interesting element of human-AI interactions that is not clearly captured by Grice’s Cooperative Principle. This can be seen in the following exchange:
 
  \ex. \label{ex:timer}
\textbf{User:} Alexa, set a timer for—\\
\textbf{Alexa:} [interrupting] Timer for how long? \textit{[rating: 4]}\\
\textbf{User:} Ugh, that’s so annoying.\\
 
In broad terms, participants did not have a problem interrupting or speaking over Alexa. However, participants were particularly averse to Alexa interrupting them, as in the above exchange. In other words, many participants demonstrated a strong assumption the human interlocutor begins every exchange, or has the floor, in any given human-AI interaction. This observation was supported by similar situations, such as when Alexa makes a suggestion seemingly unrelated to the previous exchange. This can be seen here:
 
 \ex. \label{ex:cranky}
\textbf{User:} Alexa, what’s the weather today? \\
\textbf{Alexa:} In Seattle, it’s 64 degrees Fahrenheit with mostly cloudy skies. Today, you can look for cloudy skies, with a high of 66 degrees and a low of 52 degrees. By the way, there are more things to try. Just say ‘what are your top skills?’ \textit{[rating: 5]}\\
\textbf{User:} I don’t know, I don’t always love the extra tips, but that’s me being cranky.\\
 
Here, the participant explicitly mentions her frustration at Alexa offering a seemingly unrelated request. In other words, the participant is demonstrating a dispreference for Alexa initiating a turn (though it should be noted that not all participants share this dispreference for suggestions).
 
These observations suggest that humans inherently take priority over AI in human-AI interactions. In this way, it is reminiscent of Harris’ (\citeyear{harris_1995}) notion of a “unequal encounters.” However, previous researchers have examined the extent to which context influences the speech acts available or unavailable to two human interlocutors. In the case of human-AI interactions, the inequity of the encounter is inherent to the AI and does not vary by context. In other words, there is an apparent power difference at play between the interlocutors, but it is intrinsic to the AI, rather than the context.
 
For this reason, coupled with the observations from this study, we hypothesize that another maxim — a Priority maxim — is at play in human-AI interactions. The Priority maxim can be summarized as follows: Humans take conversational priority over AI. This priority is driven by the inherent power imbalance between humans and AI. 
 
Verification of the Priority maxim would have significant implications. First, it would demonstrate the extent to which human interlocutors accommodate conversational exchanges, even when the exchange itself challenges the prototypical notion of a ‘conversation.’ Second, the Priority maxim could have important implications in terms of variation across interlocutors. For example, it may be the case that younger, more experienced AI users adhere the Priority maxim much more strongly than older, less experienced AI users. Furthermore, the Priority maxim may be implicitly acquired much like Grice’s four maxims, and the conditions under which the maxim are acquired would be important for future work. Lastly, the Priority maxim can be important guiding principle in the design of AI. The Priority maxim suggests that power dynamics are at play in human-AI interactions at a turn-taking level, where humans are assumed to take priority in initiating turns. Developing proper turn-taking strategies is notoriously difficult in AI design \citep{portela_granell-canut_2017, liao_etal_2016, chaves_gerosa_2018}. Taking the Priority maxim into account could help resolve some of these issues, where designers can take human expectations of their priority into account.

\subsection{Limitations}

Of course, there are limitations in the application of a framework meant to describe human conversation to a novel form of interaction. First, it is not entirely clear whether people perceive AI to be human-like enough to apply conversational rules at all. Second, given that exchanges with AI are currently largely transactional (e.g., the human has a desired goal, and the AI is expected to fulfill this desired goal), one could argue that an exchange between a human and AI is not necessarily a conversation. These issues can be summarized into two broad categories: human perception of AI entity and human perception of the human-AI exchange.
 
With regard to human perception of AI, it is clear that AI has no free will to actually follow a maxim. Nevertheless, several participants in the study reported here personified the device, even applying verbs of volition to describe the AI’s performance. For example, participants described Alexa’s failure to perform a particular action with language like “She tried” and “She did the very best that she could.” Such descriptions indicate that participants — whether driven by the wake word, the marketing of the product, or the fact that voice assistants are specifically designed to be human-like and conversational — speak about AI as a volitional entity. It is thus a natural step for them to also apply Grice’s maxims to such an entity. 
 
With regard to human perception of the human-AI exchange, several participants in the study also explicitly mentioned that they believe human-AI interactions to be cooperative: they were both willing and eager to cooperate with the AI, placing blame on themselves for unsuccessful exchanges or considering the exchange to be game-like. One participant blamed himself for an unsatisfactory Alexa response, saying, “I should’ve been more specific with my question. That’s on me.” Another participant, after several attempts to rephrase a question, expressed enjoyment rather than frustration, saying, “It’s actually part of the fun, I figure out oh, she needs me to speak like this.” In this way, participants demonstrated the they consider the exchange to be cooperative, not simply one-sided in favor of the human. Given that the exchange is language-based, it is therefore natural that they would also assume the conversation to be guided by Grice’s maxims.
 
Lastly, for both perception of AI and the exchange, it should also be noted that machine learning technology is rapidly evolving, making voice assistants more human-like and conversational. In this way, we can expect human perception of the AI’s conversational role to evolve alongside technological advances.

\section{Conclusion}

This study has demonstrated that Grice’s Cooperative Principle is a useful framework for understanding human assessments of their interaction with AIs. Based on the in-lab observations of twenty-three participants interacting with an Alexa device, we report on over 700 exchanges that we identified as potential maxim violations by the AI. Exploring participant commentary referencing the potential maxim violations shows that, in broad terms, participants do seem to be guided by these maxims in their evaluation of interactions with AI; they judge interactions similarly to human-human interactions, but with some small differences. 
 
Specifically, the computer component of entities like Alexa seems to play an important role in terms of user perceptions of AI. With Quantity, users do not seem to be bothered by what is arguably ‘too much’ information, and this expectation is likely driven by user expectations of a machine. In a similar vein, for Quality, users are extremely trustful of Alexa’s responses. This seems to be driven by assumptions that a computer query will deliver accurate results. Of course, this may or may not be correct assumption, and it is an important phenomenon that requires further research. 
 
In addition, this study shows that Relevance is the most salient maxim violation for the in-lab participants, occurring most frequently and rated lower on average than the other maxim violations. In other words, users seem to be more keyed in to violations of Relevance than any of the other maxims. This observation is particularly useful for the development of future AIs, where relevant exchanges seem to be key to user perceptions.
 
We also observe that there is an inherent power difference between humans and AI, where humans take priority in all exchanges, as evident in turn-initiating sequences and interruptions. To accommodate these observations, we posit a Priority maxim that guides human-AI interactions. Of course, further research needs to be done to validate the Priority maxim, as well as the extent to which Grice’s Cooperative Principle helps explain user perceptions of human-AI interactions. 
 
This work is critical for multiple reasons. First, from an AI development perspective, in order to improve AI systems like Alexa, it is important to know how users assess exchanges and what framework they are using to make those judgments. Second, while human-AI interactions may stretch the boundary of the prototypical ‘conversation,’ it is clear that human-AI interactions are becoming more and more prolific. There are multiple voice assistants available, and the contexts in which they are available is ever-increasing. This all means that the volume and variety of language-based interactions in which humans are engaging with AI is steadily increasing, and, in our opinion, this novel use of language warrants further research. 

\setlength{\bibsep}{0pt plus 0.3ex}
\setlength{\bibhang}{0.3in}			
\titleformat{\section}{\normalfont\bfseries}{\thesection}{.5em}{}		

\bibliographystyle{sp.bst}		
\newcommand{\doi}[1]{\href{http://dx.doi.org/#1}{http://dx.doi.org/#1}}	
\bibliography{lsa_bib.bib}			

\end{document}